\documentclass{PoS}
\pdfoutput=1

\usepackage[utf8]{inputenc}
\usepackage{amsmath}

\title{Searches for new heavy particles coupling to third-generation quarks at CMS}

\ShortTitle{Searches for new heavy particles coupling to third-generation quarks at CMS}

\author{\speaker{Arne Christoph Reimers}\\
		on behalf of the CMS Collaboration\\
        Universit\"at Hamburg\\
        E-mail: \email{Arne.Reimers@cern.ch}}

\abstract{Results from searches for new particles with enhanced couplings to third-generation quarks are presented. They are based on proton-proton collision data at a center-of-mass energy of 13\,TeV recorded by the CMS experiment. The signatures include single and pair production of vector-like quarks and heavy resonances decaying to third-generation quarks. A wide range of final states, from multi-leptonic to entirely hadronic is covered. Jet substructure techniques are employed to identify highly boosted heavy standard model particles in their hadronic decay modes.}

\FullConference{
European Physical Society Conference on High Energy Physics - EPS-HEP2019 -\\
			10-17 July, 2019\\
			Ghent, Belgium}

\begin{document}

\section{Introduction}

With the discovery of the Higgs boson (H) with a mass of about 125\,GeV by the ATLAS and CMS Collaborations in 2012~\cite{Aad:2012tfa, Chatrchyan:2012xdj}, the particle content of the standard model (SM) has been completed. However, the origin of its mass at the electroweak scale remains unknown. The radiative corrections to the bare mass of the Higgs boson scale quadratically with the energy scale. Hence, the individual contributions of the SM particles require an unnatural amount of fine-tuning in order to cancel and yield the measured Higgs boson mass at higher scales.

Many theories beyond the SM address this hierarchy problem by postulating the existence of new particles, like heavy spin-1 resonances~\cite{Agashe:2006hk} or non-chiral fourth generation of quarks~\cite{Marzocca:2012zn}. The left- and right-handed components of these quarks transform in the same way under the electroweak gauge group are called \textit{vector-like quarks} (VLQs) and are not excluded by measurements of Higgs boson properties. In some of these models~\cite{Greco:2014aza, Bini:2011zb}, heavy vector bosons and VLQs appear simultaneously. In order to stabilize the Higgs boson mass, the new particles are predicted to cancel the largest contributions to the correction of the Higgs boson mass, which originate from top quark (t) loops. Consequently, a large coupling to third-generation quarks is needed.

\section{Recent results from CMS}

Here, selected results of CMS searches for VLQs and heavy vector bosons obtained recently are presented. All analyses are based on the dataset collected by the CMS experiment~\cite{Chatrchyan:2008aa} in proton-proton collisions of $\sqrt{s}=13\,$TeV in the year 2016, which corresponds to an integrated luminosity of 35.9\,fb$^{-1}$. 

In all searches presented here, highly Lorentz-boosted heavy particles are expected in the final state. The boost arises from the potentially large mass difference between the new particle and its decay products. The subsequent hadronic decay products are often reconstructed in a single jet with large distance parameter ($R$) and a distinct substructure. Algorithms to identify (tag) such jets are crucial tools and widely used in CMS. The jet mass after subtracting soft and wide-angle radiation is a variable sensitive to the origin of the jet. In addition, ratios of the \textit{N-subjettiness} variable~\cite{Thaler:2010tr} are utilized. Tagging subjets as originating from b quark decays provides additional discrimination power for jets produced from decays of top quarks or Higgs bosons.

\subsection{Search for a heavy spin-1 resonance decaying to t$\overline{\text{t}}$}

The CMS search for a heavy spin-1 resonance (Z$^\prime$) decaying to a t$\overline{\text{t}}$ pair~\cite{Sirunyan:2018ryr} has been performed in all three possible decay modes of the t$\overline{\text{t}}$ system. A full statistical combination of the individual channels maximizes the sensitivity of this analysis. 

\begin{figure}[t]
\centering
\hspace{5pt}
\includegraphics[width=0.455\textwidth]{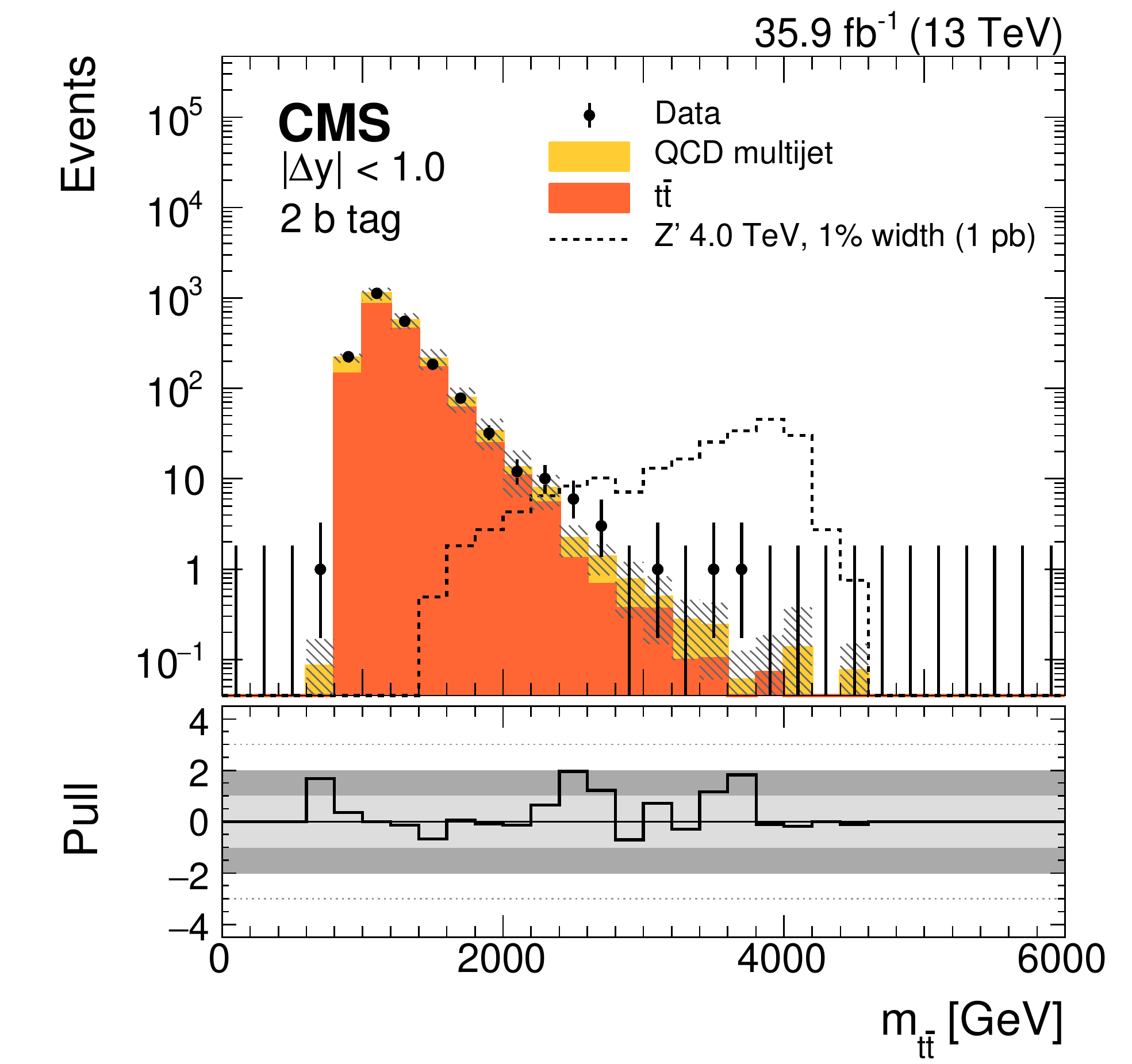}
\hfill
\includegraphics[width=0.455\textwidth]{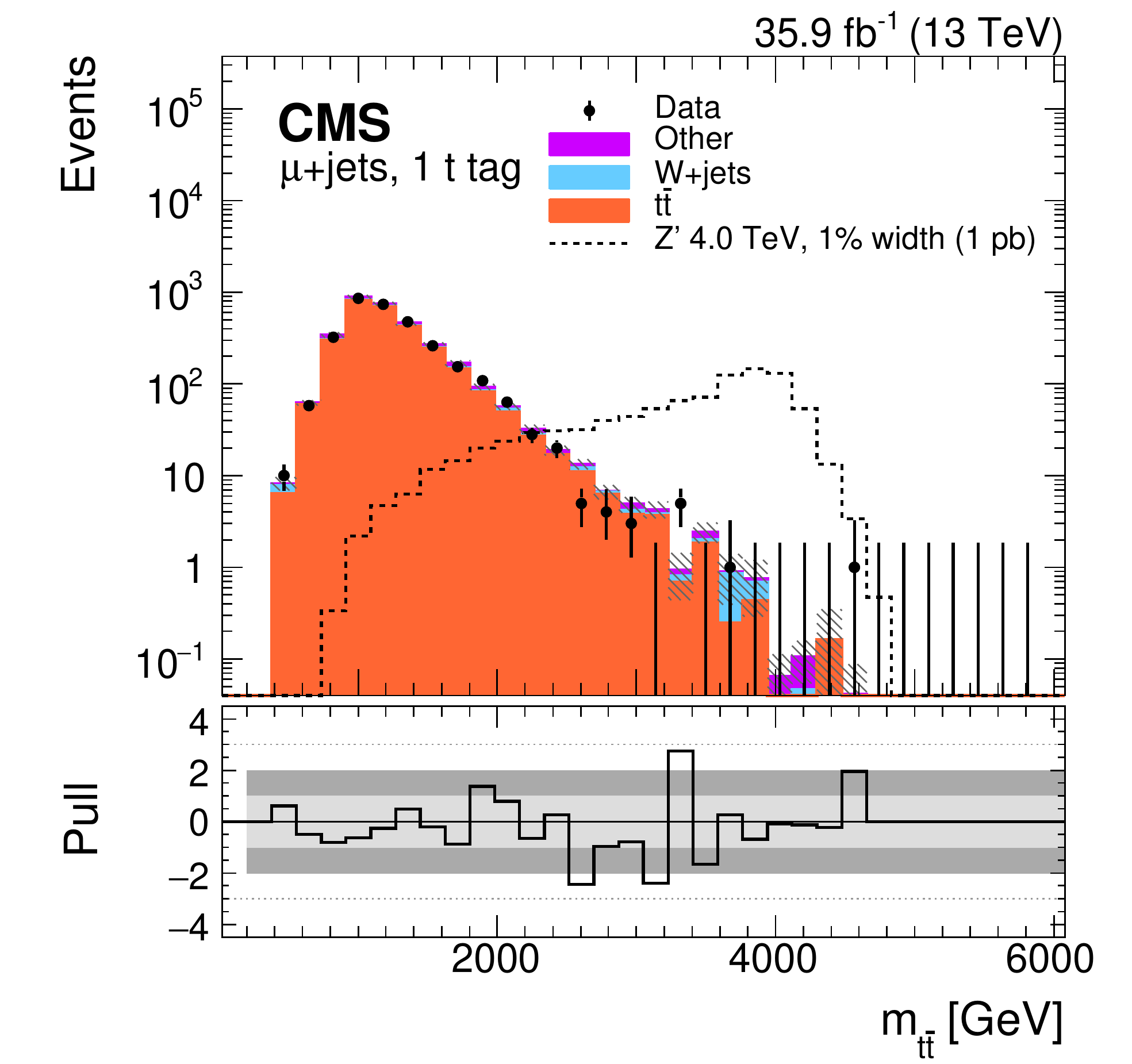}
\hspace{5pt}
\caption{Left: Distribution of $m_{\text{t}\overline{\text{t}}}$ in the all-hadronic channel in the most sensitive category. Right: Distribution of $m_{\text{t}\overline{\text{t}}}$ in the $\ell$+jets channel in events with a muon and at least one t-tagged jet. Taken from Ref.~\cite{Sirunyan:2018ryr}.
}
\label{fig:Mttbar}
\end{figure}

The all-hadronic channel is analyzed using events with at least two large-$R$ jets in the central region of the detector that exhibit a back-to-back topology in the azimuthal angle. Requiring at least two such jets to be t-tagged provides good discrimination power against events from quantum chromodynamics (QCD) multijet production. Six categories of events are defined by the rapidity difference of the two t-tagged jets with the highest $p_{\text{T}}$ ($\left| \Delta y \right| < 1.0$ or $\left| \Delta y \right| > 1.0$) and the total number of subjet b-tags (0, 1, or 2). The t$\overline{\text{t}}$ system is reconstructed using the two $p_{\text{T}}$-leading t-tagged jets and its invariant mass $m_{\text{t}\overline{\text{t}}}$ is the variable sensitive to the presence of a potential signal. The distribution of $m_{\text{t}\overline{\text{t}}}$ is shown in figure~\ref{fig:Mttbar} (left) in the most sensitive category.

The $\ell$+jets channel is enriched in signal events requiring exactly one muon or electron and at least two small-$R$ jets. The contribution of events from W+jets production is suppressed by applying a boosted decision tree trained on jet-related quantities. Events are categorized based on the presence of a t-tagged jet. The t$\overline{\text{t}}$ system is reconstructed by associating the missing transverse momentum, the lepton, and at least one jet to the top quark decaying leptonically. The hadronic top quark decay is reconstructed either from a t-tagged jet or a combination of small-$R$ jets. The distribution of $m_{\text{t}\overline{\text{t}}}$ is again used as the sensitive variable in this channel. It is shown in the muon channel for the events with at least one t-tagged jet in figure~\ref{fig:Mttbar} (right).

The dilepton channel is analyzed in events with at least two small-$R$ jets and two electrons, two muons, or one electron and one muon. At least one of the two $p_{\text{T}}$-leading jets is required to be b-tagged in addition. Both leptons are expected to be in proximity to 0 jets in signal events depending on the boost of the top quarks they originate from. The sum of distances in $R$ of the two leptons to the closest jet is therefore used to define a boosted and a non-boosted category of events for each of the three dilepton channels analyzed. The variable $S_{\text{T}}$, which is defined as the scalar sum of the transverse momenta of all selected jets and leptons and the missing transverse energy, is used to quantify the presence of a potential signal. 

\begin{figure}[t]
\centering
\hspace{5pt}
\includegraphics[width=0.50\textwidth]{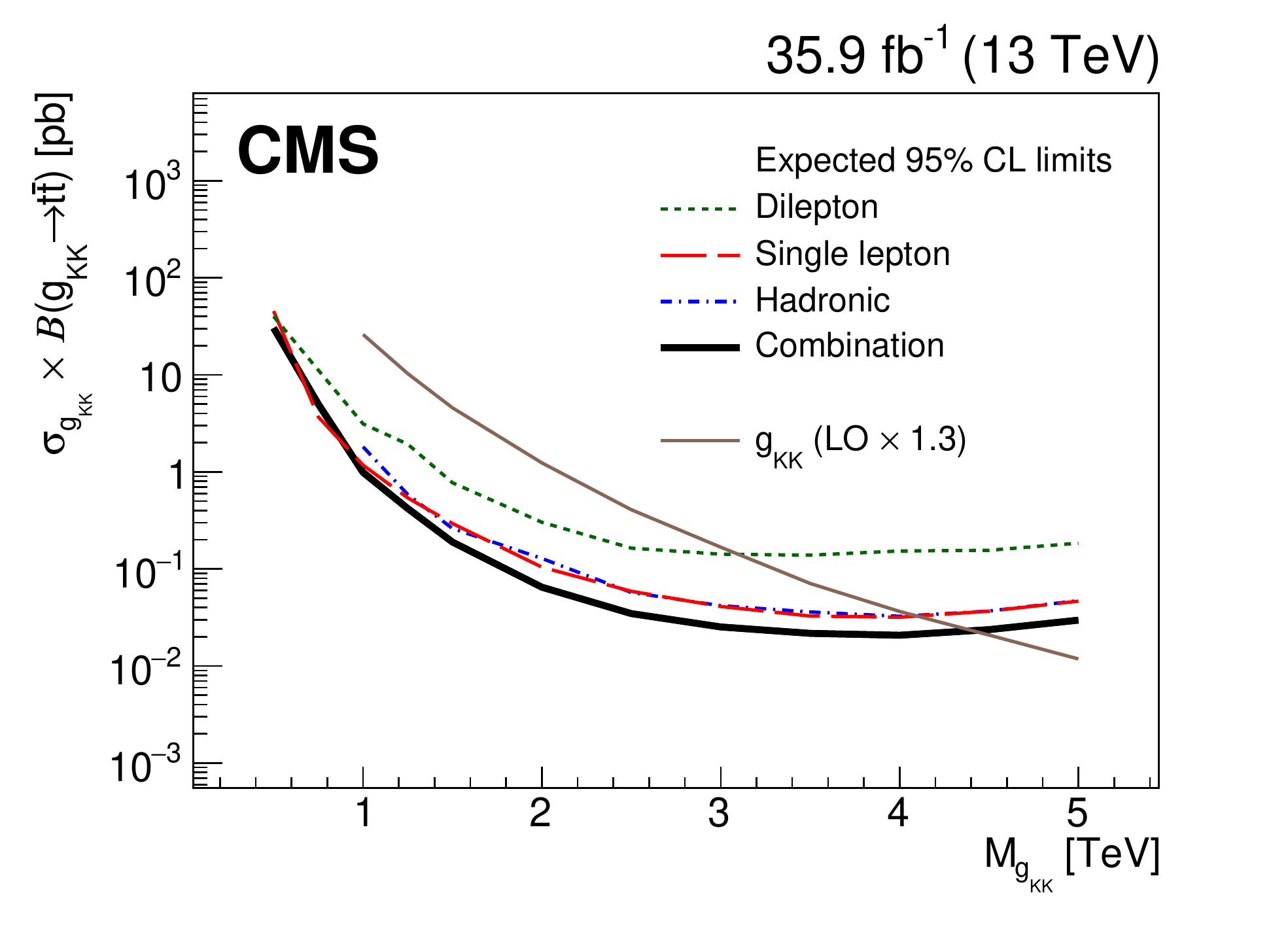}
\hfill
\includegraphics[width=0.455\textwidth]{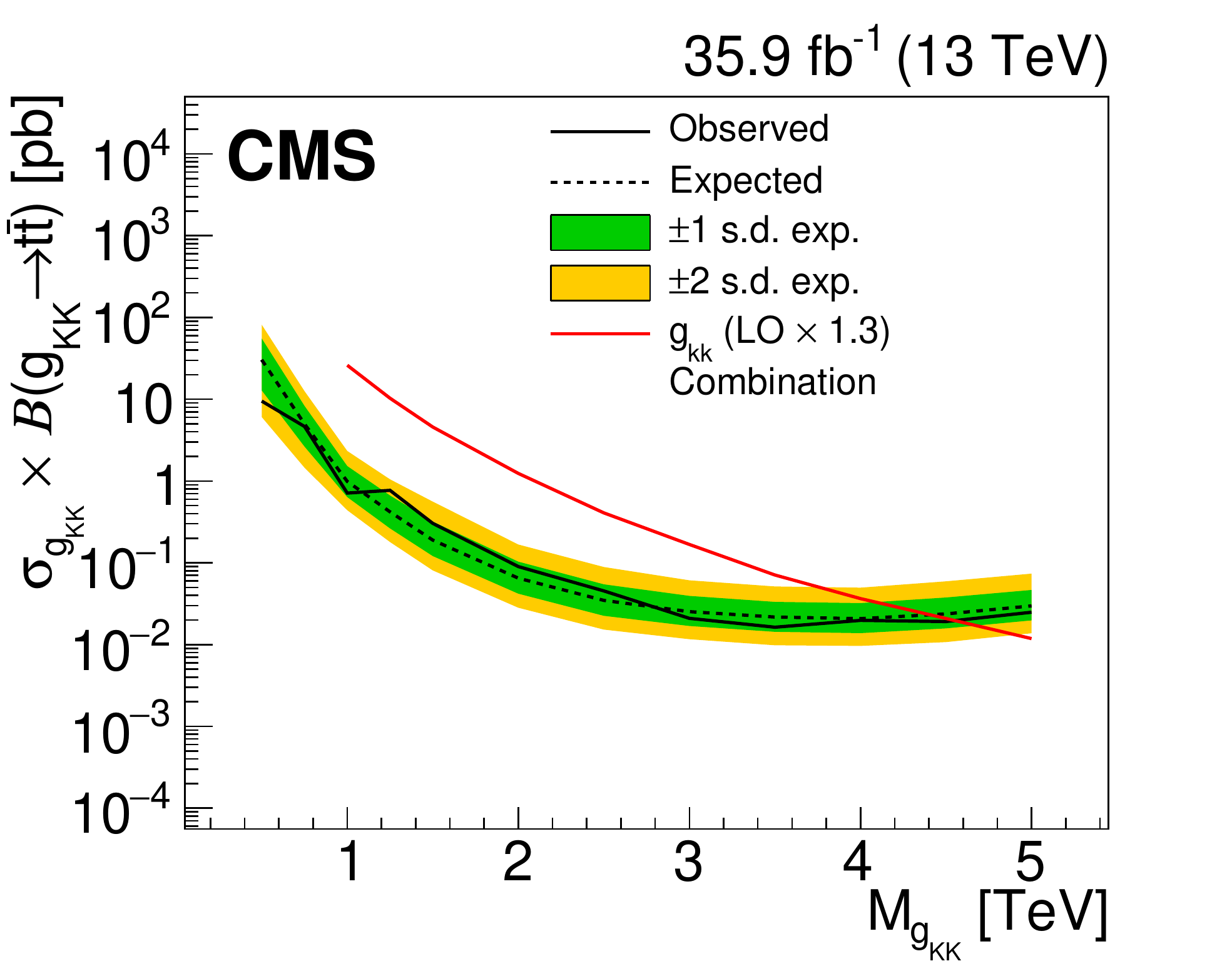}
\hspace{5pt}
\caption{Left: Expected cross section upper limits on the production of Kaluza-Klein excitations of the gluon decaying exclusively to a pair of top quarks. The contributions of the individual limits and the combined result are shown. Right: Combined observed and expected cross section upper limits on the production of Kaluza-Klein excitations of the gluon decaying exclusively to a pair of top quarks. Taken from Ref.~\cite{Sirunyan:2018ryr}.}
\label{fig:TTbarLimits}
\end{figure}

The analyses conducted in the three possible final states of the t$\overline{\text{t}}$ decay are combined by performing a simultaneous fit to the distributions of sensitive variables introduced above. In all cases, no deviation from the SM expectation is observed and upper limits on the production cross section of new spin-1 resonances are placed. The contributions of the individual channels and the combined expected result are presented in figure~\ref{fig:TTbarLimits} (left) while expected and observed limits are compared in figure~\ref{fig:TTbarLimits} (right). Kaluza-Klein excitations of the gluon are excluded up to masses of 4.55\,TeV under the assumption of unit branching fraction for the decay to a pair of top quarks.

\subsection{Search for a heavy spin-1 resonance decaying to a top quark and a vector-like top quark}

Since some models beyond the SM predict heavy vector resonances that are accompanied by VLQs, the CMS collaboration has conducted a search for a new Z$^\prime$ boson decaying to a top quark and a heavy vector-like top partner (T)~\cite{Sirunyan:2018rfo}. This decay mode is dominant for intermediate Z$^\prime$ masses with $M_{\text{t}} + M_{\text{T}} < M_{\text{Z}^\prime} < 2 M_{\text{T}}$. 

This analysis is optimized for subsequent decays of the T quark to tH or tZ. It is carried out in the final state with exactly one electron or muon, which is assumed to originate from the decay of one of the top quarks, and at least one large-$R$ jet. At least one such jet is required to be tagged as originating from the decay of an H or W/Z boson. Additionally considering the multiplicity of subjet b-tags (1 or 2) in the H-tag category, events are sorted into three exclusive categories by this requirement. Each category is then divided into events with and without an additional t-tagged jet, referred to as boosted and resolved categories. Differences in the efficiencies and mistag rates of the tagging algorithms between data and simulation are corrected with scale factors measured in dedicated control regions. The dominating t$\overline{\text{t}}$ and W+jets backgrounds are constrained in control regions without any H- or W/Z-tagged jets.

\begin{figure}[t]
\centering
\hspace{5pt}
\includegraphics[width=0.44\textwidth]{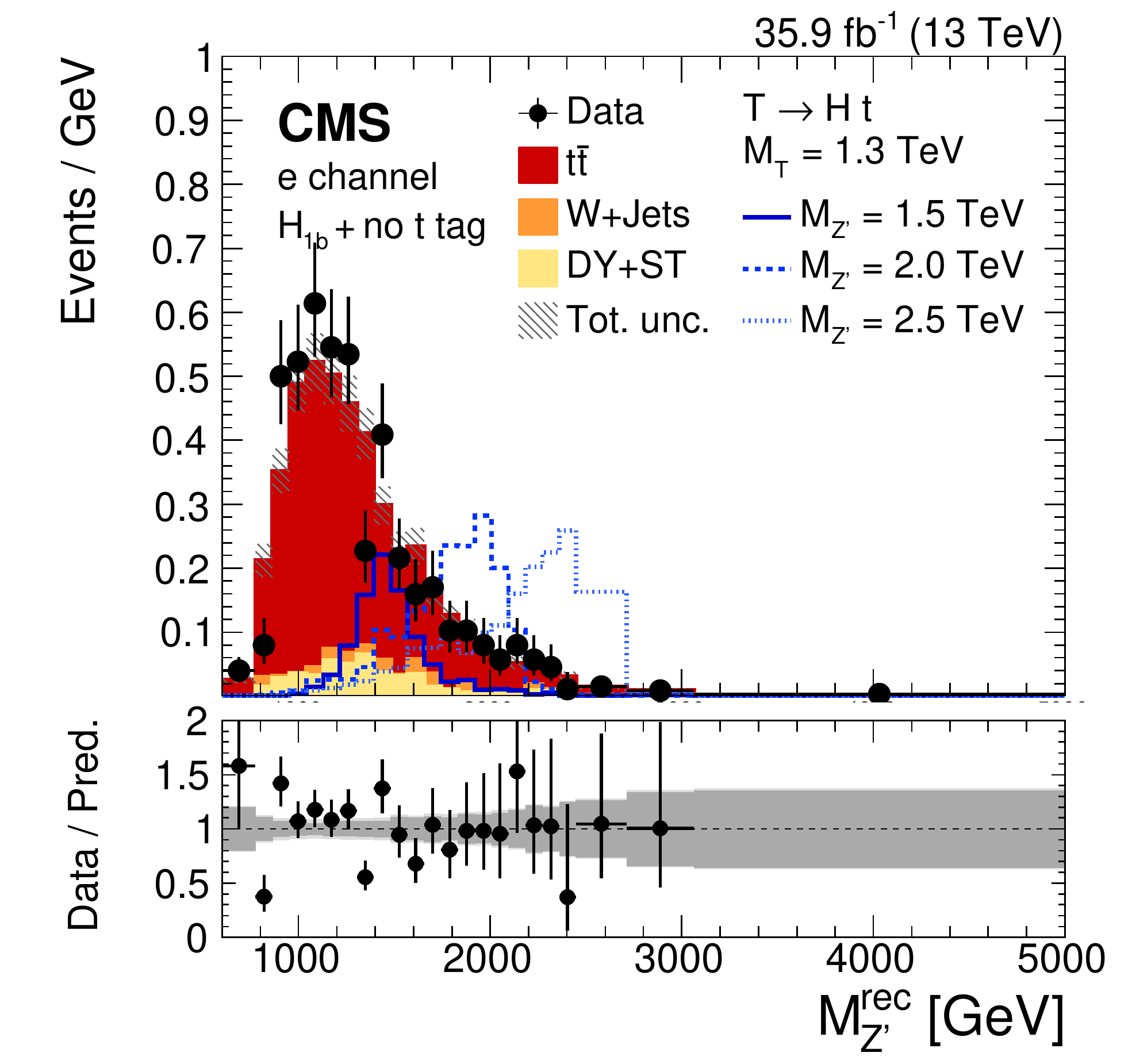}
\hfill
\includegraphics[width=0.44\textwidth]{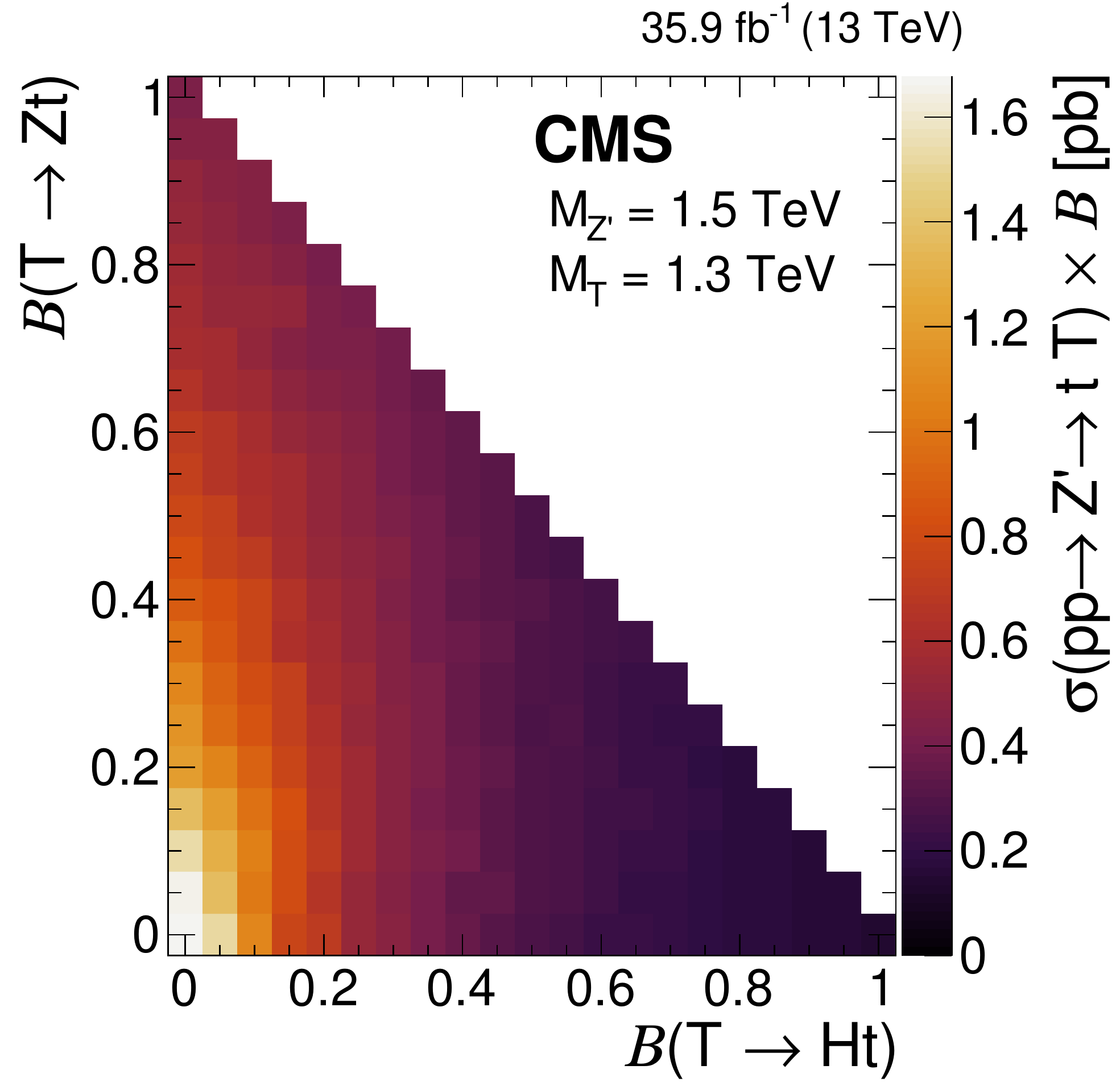}
\hspace{5pt}
\caption{Left: Distribution of $M_{\text{Z}^\prime}^{\text{rec}}$ in the category with one electron, a H-tagged jet with one subjet b-tag and no additional t-tagged jet. Right: Observed cross section upper limits on the Z$^\prime$ production for $M_{\text{Z}^{\prime}} = 1.5$\,TeV and $M_{\text{T}} = 1.3\,$TeV for varying branching fractions of the T quark decay. The lower left corner corresponds to unit branching fraction for the decay $\text{T}\to\text{bW}$. Taken from Ref.~\cite{Sirunyan:2018rfo}.}
\label{fig:ZprimeTt}
\end{figure}

The Z$^\prime$ resonance is reconstructed as follows. The Z or H boson is reconstructed with the jet tagged correspondingly. The top quark decaying to the hadronic final state is reconstructed from the t-tagged jet in the boosted categories and from a combination of small-$R$ jets in the resolved final states. Last, the top quark decaying to the leptonic final state is reconstructed from the electron or muon, missing transverse energy, and small-$R$ jets. The reconstructed Z$^\prime$ mass, $M_{\text{Z}^\prime}^{\text{rec}}$, is used in the final statistical evaluation to quantify the presence of a potential signal. The distribution of $M_{\text{Z}^\prime}^{\text{rec}}$ in one category is shown in figure~\ref{fig:ZprimeTt} (left).

No excess over the SM expectation is observed and upper limits on the production cross section of new Z$^\prime$ bosons are placed. They depend on the masses of the Z$^\prime$ boson and the T quark and the branching fractions for the T quark decay, where the tZ, tH, and bW decay modes are considered. For $M_{\text{Z}^{\prime}} = 1.5$\,TeV and $M_{\text{T}} = 1.3\,$TeV, the observed upper limits are shown figure~\ref{fig:ZprimeTt} (right) for varying branching fractions. They are the most stringent in the $\text{Z}^\prime\to\text{Tt}\to\text{tHt}$ channel to date.

\subsection{Search for heavy partners of the top or bottom quark in the all-hadronic final state}

The CMS search for heavy partners of the t or b quark (B) in the all-hadronic final state~\cite{Sirunyan:2019sza} considers three possible decay modes of the T (B), $\text{T}\to \text{tH},~\text{tZ},~\text{bW}$ ($\text{B}\to \text{bH},~\text{bZ},~\text{tW}$). The analysis utilizes the Boosted Event Shape Tagger (BEST)~\cite{PhysRevD.94.094027}, a neural-network-based algorithm optimized to identify jets originating from boosted heavy particles. It makes use of the distinct topology that the decay products of a heavy particle exhibit in its rest frame. After boosting the jet constituents into different rest frames, event shape variables serve as input to the neural network and the output class with the highest score defines the tag of a jet. Jets are associated to one out of six possible classes, which correspond to t, H, W, Z, b, or light jets.

\begin{figure}[t]
\centering
\hspace{5pt}
\includegraphics[width=0.44\textwidth]{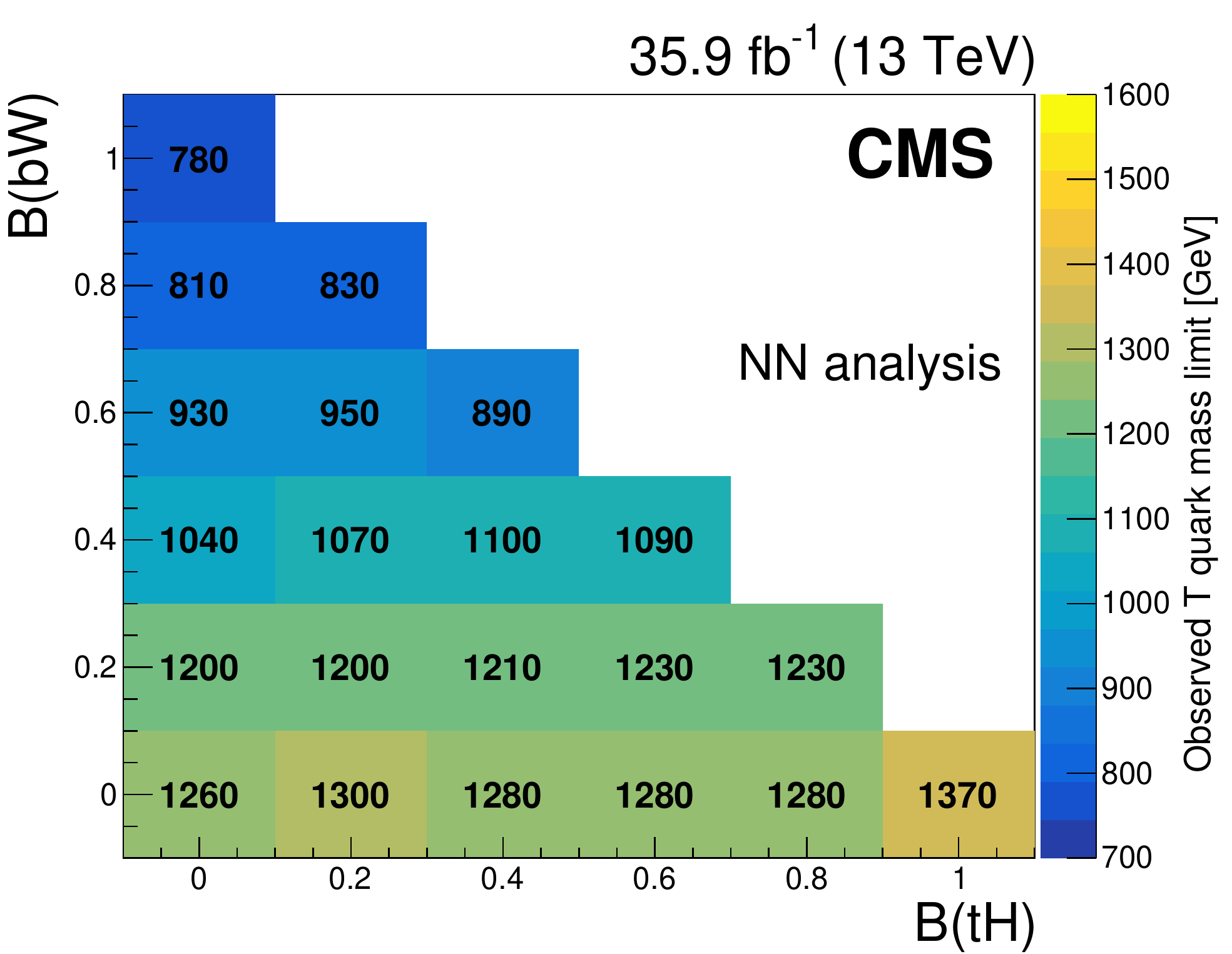}
\hfill
\includegraphics[width=0.44\textwidth]{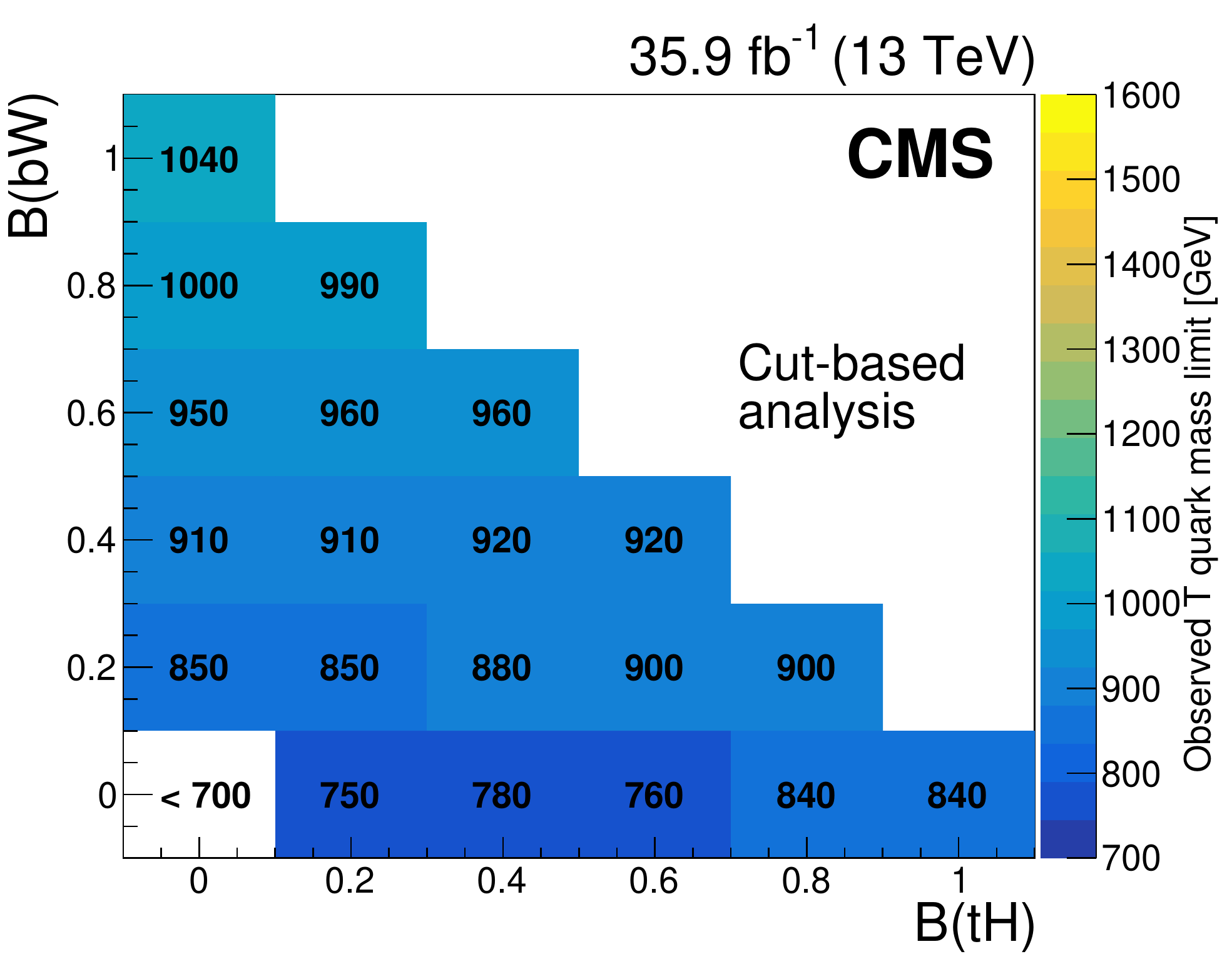}
\hspace{5pt}
\caption{Observed upper limits on the mass of pair-produced T quarks for varying branching fractions of the T quark decay obtained with the BEST-based analysis (left) and the cut-based analysis (right). The bottom left corner in each plot corresponds to a branching fraction of one for the decay $\text{T}\to\text{tZ}$. Taken from Ref.~\cite{Sirunyan:2019sza}.}
\label{fig:BESTLimits}
\end{figure}

Events with exactly four large-R jets are considered. The 0 of jets tagged in different BEST categories are used to define 126 exclusive event categories. 
The summed energy of the four jets, $H_{\text{T}}$, is used as a sensitive variable to discriminate SM backgrounds from signal in the final statistical analysis. If the number of expected events is too small in a given category, a single bin is used instead. The BEST-based analysis performs best in final states with a high multiplicity of boosted heavy particles. Therefore, it is complemented by a cut-based analysis optimized for the $\text{TT}\to\text{bWbW}$ final state using the $H_{\text{T}}$ variable computed from small-$R$ jets as a sensitive variable. No excess over the SM expectation is observed and upper limits on the pair production cross section of T and B quarks are placed for varying branching fractions for the three decay modes considered. The mass exclusion limits are presented in figure~\ref{fig:BESTLimits} for the BEST-based analysis (left) and for the cut-based analysis (right) for T quark pair production. T quark masses below 950\,GeV are excluded for all possible branching fractions. These results represent the most stringent limits on T and B quark pair production in the all-hadronic decay channel and are comparable to those obtained from searches in leptonic final states.

\subsection{Search for single production of VLQs in the tW decay channel}

The CMS analysis searching for single production of VLQs decaying to a top quark and a W boson~\cite{Sirunyan:2019tib} is performed in the final state with exactly one electron or muon, missing transverse energy, and several jets. This signature can be produced by VLQs of different charges, i.e. a B quark with a charge of $-1/3$\,e or a VLQ with charge $+5/3$\,e ($\text{X}_{5/3}$). The single production of VLQs is expected to be relevant especially at high VLQ masses, for which the pair production mode is kinematically suppressed.

The electron or muon in the event is always expected to originate from the decay of a W boson, associated by a neutrino. Therefore, this W boson is reconstructed from the missing transverse energy and the lepton reconstructed in the event. If a t-tagged jet is identified, it is taken as the candidate for a top quark decaying to the all-hadronic final state. Otherwise, a combination of small-$R$ jets is used to reconstruct the hadronic part of the VLQ decay. Finally, the reconstructed VLQ mass, denoted $m_{\text{reco}}$, is defined as the invariant mass of the leptonic W candidate and the hadronic part.

Events are split into exclusive categories based on the presence of a t-, W-, or b-tagged jet. In events of single VLQ production, an additional jet is expected to be emitted under a low angle to the beam axis. It is reconstructed in the region of high $\left| \eta \right|$ in the detector and its presence is exploited to define signal- or background-enriched data samples. In the central control regions, the background composition and shape is similar to the signal-enriched regions that contain a jet reconstructed at high $\left| \eta \right|$. The full background prediction in all signal regions is obtained from data in the control regions corrected for residual kinematic differences smaller than 20\%. 

\begin{figure}[t]
\centering
\includegraphics[width=0.39\textwidth]{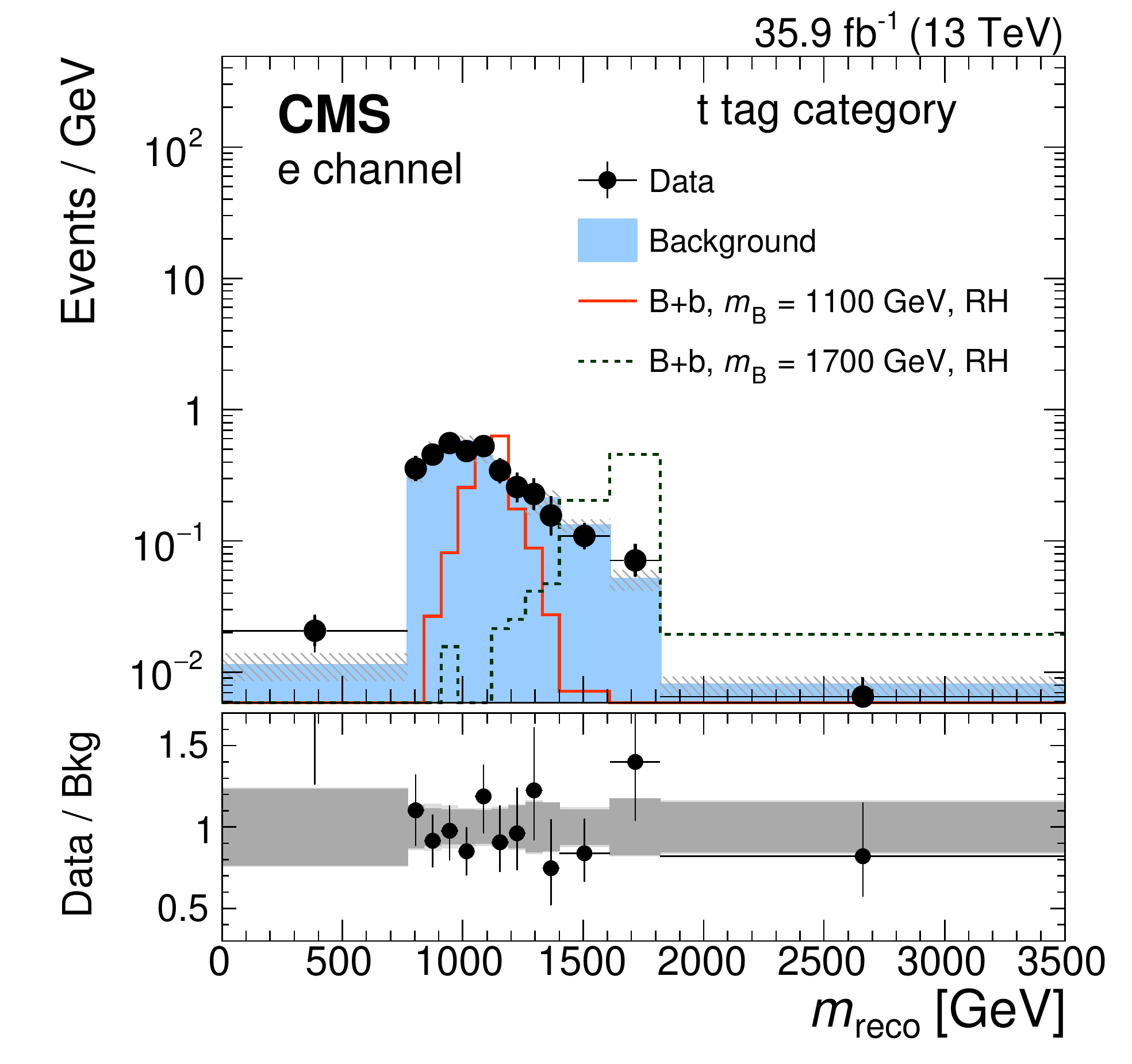}
\hfill
\includegraphics[width=0.53\textwidth]{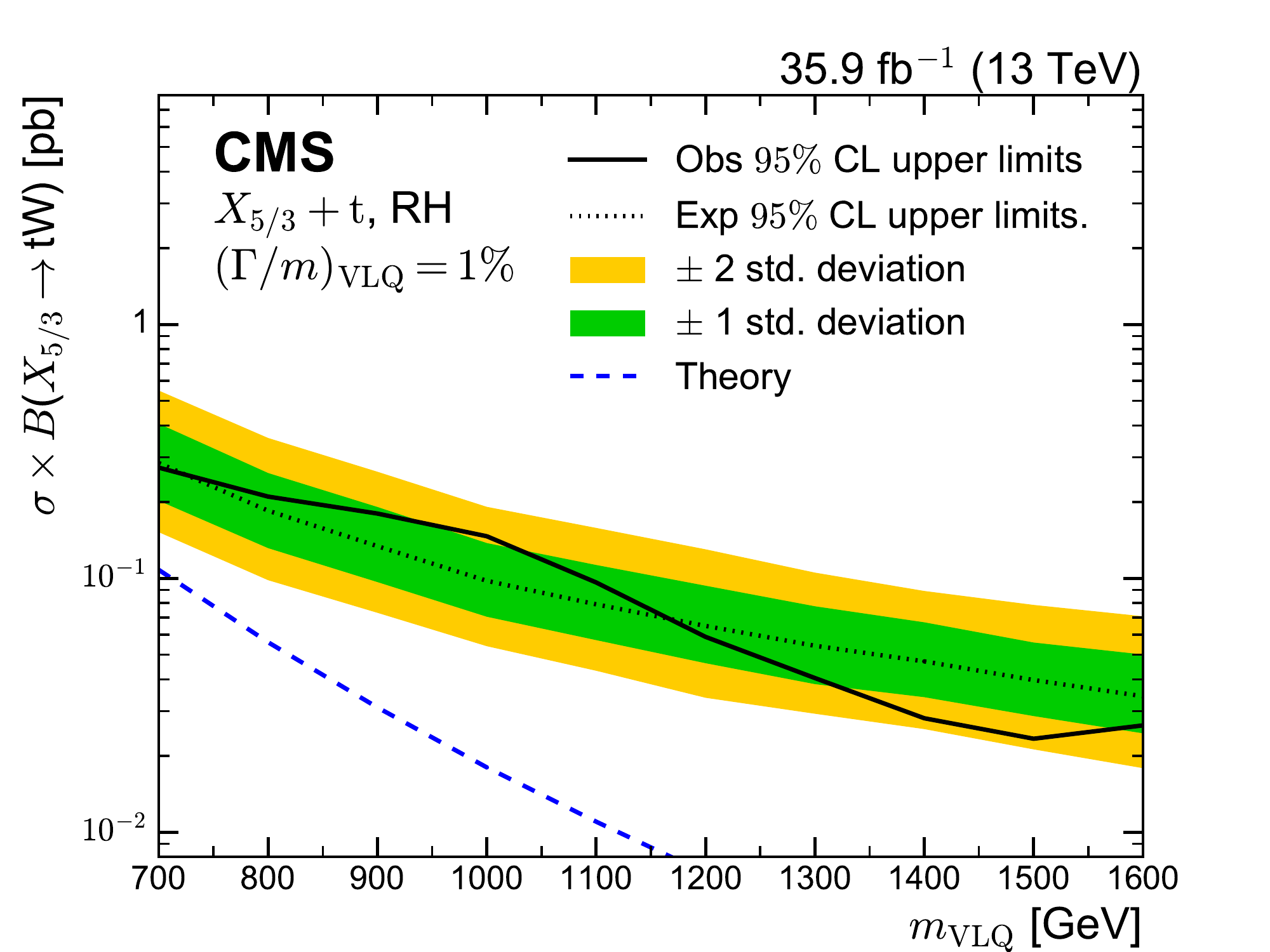}
\caption{Left: Distribution of $m_{\text{reco}}$ in the category with a t-tagged jet in the e+jets channel with a fully data-drive background estimation. Right: Observed and expected cross section upper limits on the single production of an $\text{X}_{5/3}$ quark decaying exclusively to a top quark and a W boson. Taken from Ref.~\cite{Sirunyan:2019tib}.}
\label{fig:SingleVLQ}
\end{figure}

The distribution of $m_{\text{reco}}$ in one category is shown in figure~\ref{fig:SingleVLQ} (left). Good agreement between data and the data-driven SM expectation is observed throughout all categories and upper limits on the single production cross section of B and $\text{X}_{5/3}$ quarks are set, assuming they exclusively decay to a top quark and a W boson. The observed and expected limits on single production of a narrow $\text{X}_{5/3}$ quark are shown in figure~\ref{fig:SingleVLQ} (right). They constitute the most stringent constraints on single VLQ production in this channel to date.

\bibliographystyle{JHEP}
\bibliography{bibliography}

\end{document}